# The Extensive Special Relativity and Comment on Local Lorentz Transformation in Varying Speed of Light Theory

Yi-Fang Chang

*Department of Physics, Yunnan University, Kunming, 650091, China*

(E-mail: yifangchang1030@hotmail.com)

**Abstract**

It is proved that local Lorentz transformations for different systems cannot derive varying speed of light. Based on the special relativity principle, an invariant speed $c_h$ is necessarily obtained. Therefore, the exact basic principles of the special relativity should be redefined as: I. The special relativity principle, which derives necessarily an invariant speed $c_h$. II. Suppose that the invariant speed $c_h$ in the theory is the speed of light in the vacuum $c$. If the second principle does not hold, for example, the superluminal motions exist, the theory will be still the extensive special relativity, in which the formulations are the same, only c is replaced by the invariant speed $c \rightarrow c_h$. If the invariant speed $c_h$ are various invariant velocities, the diversity of space-time will correspond to many worlds. Varying speed of light theory is probably connected only with the general relativity.

Key words: special relativity, many worlds, varying speed of light, Lorentz transformations, cosmology, invariant speed

PACS: 03.30.+p, 98.80.Hw

## 1. Introduction

Recently, the varying speed of light (VSL) theory is discussed warmly [1-13]. It is applied to interpretation of horizon, flatness and cosmological constant problems. As a possible development of the general relativity, it is very meaning. Unfortunately the foundations of such theory are far from solid. Magueijo noted that nothing prevents the construction of a theory satisfying the principle of relativity while still allowing for space-time variations in the velocity of light c [10]. He proposed as long as these concepts are subject to very minimal generalization [13].

Magueijo redefines units of time and space in all inertial systems [10]:

$$d\hat{t} = dt\varepsilon^{\alpha}, d\hat{x}^i = dx^i \varepsilon^{\beta}, \tag{1}$$

where $\varepsilon$ can be any function. If $\alpha \neq \beta$, a Lorentz invariant theory is replaced by a theory in which c remains isotropic, color independent, and independent of the speeds of observer and emitter, but it varies like $\hat{c} \propto \varepsilon^{\beta-\alpha}$. From this Magueijo obtains local Lorentz transformations in the new units are preserved [10]:

$$d\hat{t}' = \gamma(d\hat{t} - \frac{\hat{v}}{\hat{c}^2}d\hat{x}), d\hat{x}' = \gamma(d\hat{x} - \hat{v}d\hat{t}), \tag{2}$$

with
$$\gamma = \frac{1}{\sqrt{1-(\hat{v}/\hat{c})^2}}. \tag{3}$$

Their forms are the same with usual Lorentz transformations replaced only c→$\hat{c}$.

## 2. Comment on local Lorentz transformations and varying speed of light

In the $\hat{x}_1 - \hat{x}_0$ plane, local Lorentz transformations between the inertial systems may become:

$$d\hat{x}_\mu ' = \alpha_{\mu\nu} d\hat{x}_\nu, \alpha_{\mu\nu} = \frac{1}{\sqrt{1-\beta^2}}\begin{pmatrix} 1 & -\beta \\ -\beta & 1 \end{pmatrix}. \tag{4}$$

Assume that $\hat{c}$ and $\hat{c}'$ are different in the two inertial systems K and K'. For $d\hat{x}-d\hat{t}$ system, the transformation factor is:

$$\alpha_{\mu\nu} = \gamma\begin{pmatrix} 1 & -\beta\hat{c} \\ -\beta/\hat{c} & 1 \end{pmatrix}. \tag{5}$$

For $d\hat{x}'-d\hat{t}'$ system,

$$\alpha_{\mu\nu}' = \gamma'\begin{pmatrix} 1 & -\beta'\hat{c}' \\ -\beta'/\hat{c}' & 1 \end{pmatrix}. \tag{6}$$

We may introduce another K''¡¯ system whose transformation factor should

$$\alpha_{\mu\nu}'' = \alpha_{\mu\nu}\alpha_{\mu\nu}' = \gamma\gamma'\begin{pmatrix} 1+\beta\hat{c}\beta'/\hat{c}' & -\beta\hat{c}-\beta'\hat{c}' \\ -\beta/\hat{c}-\beta'/\hat{c}' & 1+\beta\beta'\hat{c}'/\hat{c} \end{pmatrix}. \tag{7}$$

The form of the factor $\alpha_{\mu\nu}''$ must be the same with $\alpha_{\mu\nu}$, so $\alpha_{11}''=\alpha_{00}''$, i.e., $1+\beta\hat{c}\beta'/\hat{c}'=1+\beta\beta'\hat{c}'/\hat{c}, \therefore |\hat{c}|=|\hat{c}'|$. Therefore, the transformations derive necessarily that the speed $\hat{c}$ is invariant. Correspondingly, in flat space-time with metric $\eta_{\mu\nu}$=diag(-1,1,1,1), $ds^2 = d\hat{x}_i^2 - \hat{c}^2 d\hat{t}^2$ is invariant, and there are the Lorentz transformations with $\hat{c}$=constant. We cannot derive $c = \frac{c_0}{1+(c_0 t/R)}$. In this case, it is not clear and arouses suspicion that $\hat{c}$ may be varying and $\varepsilon$ can be any function [10]. This error is primary. Of course, if local Lorentz transformations cannot be applied for different systems, they will be also meaningless.

Further, we suppose that the special relativity principle holds among the inertial systems, and introduce the two inertial systems K and K', whose relative speed is u [14]. Then for an origin '

$$x' = (x-ut)/\theta(-u). \tag{8}$$

For another origin O

$$x = (x'+ut')/\theta(u). \tag{9}$$

Using the special relativity principle or the isotropy of space $\theta(-u) = \theta(u)$, from (8) and (9) we obtain

$$t' = \{t - [1-\theta^2(u)]\frac{x}{u}\}/\theta(u), \tag{10}$$

$$\frac{dx'}{dt'} = v' = (v-u)/\{1-[1-\theta^2(u)]\frac{v}{u}\}. \tag{11}$$

We introduce the K'' system whose speed is v relative to K, so the speed of '''' 's $v_i^-$ relative 'to From the same reason for the origin O'

$$\frac{dx''}{dt''} = -v' = (u-v)/\{1-[1-\theta^2(v)]\frac{u}{v}\}. \tag{12}$$

Equs. (11) and (12) are compared, and then we derive $[1-\theta^2(u)](v/u) = [1-\theta^2(v)](u/v)$. Using the separation of variables

$$[1-\theta^2(u)]/u^2 = [1-\theta^2(v)]/v^2 = k, \tag{13}$$

where the quantity k is a universal constant that is independent of relative speeds u and v. Because $\theta$ is dimensionless, the dimension of k is the same with $1/u^2$. Such we can assume that $k=1/c_h^2$. Since it cannot be determined that $c_h$ is speed of something by the logical way, we may only suppose that $c_h$ is an arbitrary invariable speed, so

$$\theta(u) = \sqrt{1-ku^2} = \sqrt{1-(u/c_h)^2}. \tag{14}$$

From (8) and (9) we obtain the extensive Lorentz transformation (LT) of c→$c_h$,

$$x' = \frac{x-ut}{\sqrt{1-(u^2/c_h^2)}}, t' = \frac{t-(ux/c_h^2)}{\sqrt{1-(u^2/c_h^2)}}. \tag{15}$$

The special relativity principle derives necessarily the invariance of a certain velocity, and the constancy-principle of this velocity [14]. In fact, some methods of derivation of the LT are universal, and do not require the constancy of the velocity of light [15-17]. For example, Lee and Kalotas presented that a method derives the LT by the principle of relativity alone, without resorting to the existence of a universal limiting velocity [16]. Therefore, the two basic principles of the special relativity are not independent one another, and are not complete on equality. We think that the basic principles of the special relativity should be redefined more suitably as [14]: I. The special relativity principle. II. Suppose that the invariant velocity $c_h$ in the theory is the velocity of light in the vacuum.

## 3. Extensive special relativity and many worlds

Now, various superluminal phenomena and theories have been researched widely. For example, in 2000 Mugnai, et al., observed superluminal behaviors in the propagation of localized microwaves over distances of tens of wavelengths, and all components in the spectral extension have the same propagation speed $v = c/\cos\theta$ [18]. Wang, et al., measured a very large superluminal group speed that exceeds about 310 times faster than the speed of light in a vacuum, but these light pulses do not violate causality [19].

Einstein pointed out the relativity of space and time. It should be universal, and does not depend on the particular constancy of the velocity of light. Only based on the special relativity principle or various equivalent principles, one may derive the extensive LT, and the universal velocity invariance. It is called the extensive special relativity, in which only c→$c_h$ (another invariant velocity). It includes the classical Newton's theory of $c_h = \infty$, and the special relativity of $c_h$=c. But the velocity is not infinite borne out by experiment [16].

The velocity $c_h$ is invariant, and $ds^2 = dx_\alpha^2 - c_h^2 dt^2$ is also an invariable quantity. But, $c_h$ may not be the velocity of light in the vacuum, so it is relative, and may be different values for different regions. The extensive special relativity bases only on the special relativity principle. The second principle in the special relativity is a concrete $c_h$. Even if it is shaken by some new experiments, the formulations of this theory will be able to be not also changing greatly. We cannot but wonder with admiration that the special theory of relativity is really a full beautiful and very universal theory. If the measurements have error, even the velocity of light is not constancy, it is also possible that the extensive special relativity holds still.

Further, the development of the special relativity seems to have only a direction: The special relativity principle is corrected. So is Einstein just, he obtained the general relativity, whose results include naturally that the speed of light is not invariant. Einstein said [20]: "In order to complete the definition of time we may employ the principle of the constancy of the velocity of light in a vacuum," and it will not lead to contradictions. Moreover, "in order to give physical significance to the concept of time, processes of some kind are required which enable relations to be established between different places. It is immaterial what kind of processes one chooses for such a definition of time." From this we may consider another possible direction $c_h$ not only can be the speed of light in the vacuum, but also can be other arbitrary invariant speed. In other words, the properties of space-time are possibly the variety under different conditions, and the invariant velocities are not only and unified. It is a development of the space-time theory of relativity. This diversity of space-time is called a variety of space-time systems. It is connected possibly with the many universe proposed by Everett [21], and after many worlds [22], etc. Moreover, the extensive special relativity and the variety of space-time systems will have probably the wide application in the general theory of relativity and in particle physics.

Avelino and Martins pointed out [5]: In the theory of Albrecht and Mogueijo, one cannot find

any choice of time unit in which the theory reduces to the standard one. It shows that this theory does not possess the correspondence principle. Assume that varying speed of light is possible, in particular for an evolutional universe, in which only the general relativity is suitable. If $ds^2 = d\hat{x}_i^2 - \hat{c}^2 d\hat{t}^2$ is not invariant, corresponding theories will be different generalizations of the general relativity. For example, it is connected possibly with the Brans-Dicke scalar-tensor theory, tensor-tensor theory, etc [11,23]. While Dirac's large numbers hypothesis should be the first theory on varying universal constants, in which second consequence is that the gravitational 'constant' must decrease with time, proportionally to $1/t$ [24]. Kostelecky, et al., proved that spacetime-varying coupling constants could be associated with violations of local Lorentz invariance [25].